\documentclass[10pt,journal,compsoc]{IEEEtran}
\usepackage{amsmath}
\usepackage{amsfonts}
\usepackage{amssymb}
\usepackage{algorithm}
\usepackage{algorithmic}
\usepackage{hyperref}
\usepackage{graphicx}

\begin{document}

\title{Real-Time Arbitrary Resampling for Phase-Continuous Microtuning in Audio Streams}

\author{Andreas~Falkenberg\\
QPI Laboratories, Inc.\\
Escondido, CA, US\\
andreas@qpilabs.com}

\maketitle

\begin{abstract}
This paper develops an output-driven resampling formulation for fine pitch adjustment of recorded signals. A fractional input-position increment combines the desired frequency ratio with independently specified input and output sample rates. Each output sample is evaluated from neighboring input samples, and gradual rate changes preserve the current input position. Musical interval mapping and direct sine synthesis are treated as distinct layers. The formulation separates tuning resolution, numerical accuracy, interpolation quality, and streaming correctness. Optional rational-factor representations use freely chosen approximation parameters; numerical illustrations do not define constants required by the algorithm. Linear interpolation provides a low-cost evaluation kernel but does not establish a universal signal-quality or frequency-accuracy guarantee.
\end{abstract}


\section{Introduction}
Fine retuning is useful when a recorded track must be aligned with another tuning reference or when related versions of a source are generated at slightly different pitches. Source and target reference frequencies are input parameters. The method operates on the waveform; it does not estimate pitch or identify individual notes within a recording. The treatment concerns signal generation and transformation rather than perceptual or therapeutic effects.

In this scenario, microtuning means applying a small multiplicative pitch adjustment to an entire source stream. It is not independent retuning of the notes within a polyphonic recording. A single conversion ratio acts on all spectral components. The generalized formulation accepts a target frequency directly or a logarithmic interval in cents, without requiring a particular piano-key numbering scheme or a fixed concert-pitch reference.

The formulation combines output-driven sample evaluation, fractional input positioning, and continuous rate control. Established bandlimited methods also support arbitrary evaluation times and smoothly changing rates~\cite{smith_gossett}. The distinction examined here is the low-cost two-sample evaluation kernel and its associated accuracy and bandwidth requirements, extending early on-demand sample creation strategies optimized for sub-Hertz frequency matching~\cite{falkenberg2018}, rather than a claim that arbitrary-time resampling is inherently novel.

\section{Generalized Model and Scope}
Let $F_{\text{in}}$ and $F_{\text{out}}$ denote positive input and output sample rates, and let $f_{\text{source}}$ and $f_{\text{target}}$ denote positive source and target reference frequencies. Define $p = f_{\text{target}}/f_{\text{source}}$ and $q = (F_{\text{in}}/F_{\text{out}})p$. For constant $q$, output sample $n$ is evaluated at $u[n] = u + nq$. For changing $q$, accumulate $u[n+1] = u[n] + q[n]$. Coordinates are measured in input-sample intervals, while the output clock remains $F_{\text{out}}$. These relationships define the algorithm independently of any programming language, block size, or fixed numerical example.

With $p = 1$, the method performs sample-rate conversion without an intended pitch change. With $F_{\text{in}} = F_{\text{out}}$, $q = p$ and playback speed and pitch change together. For constant $p$, the continuous-time interpretation is $y(t) = x(pt)$; a finite source duration $T$ becomes approximately $T/p$, apart from boundary handling. Duration-preserving pitch adjustment requires an additional process. Symbolic relationships define the method; any particular numerical substitutions in the displayed expressions are illustrations only.

\section{Microtuning: Frequency Ratios and Cents}
Microtuning can be specified directly by $p = f_{\text{target}}/f_{\text{source}}$. An interval of $m$ equal-tempered semitones gives $p = 2^{m/12}$; an interval of $c$ cents gives $p = 2^{c/1200}$, with negative $c$ lowering pitch. These parameterizations do not restrict the resampler to a particular note reference or to whole-cent steps.
\begin{equation}
f_{n+m} = f_n \cdot 2^{\frac{m}{12}}
\end{equation}

A cent is one hundredth of an equal-tempered semitone; there are 1200 cents per octave. Conversely, $c = 1200 \log_2(p)$~\cite{ellis}. The values 100 and 1200 arise from the definition of this musical unit. Unlike an arbitrarily selected rational denominator, they are not tuning-resolution parameters chosen for the resampling algorithm.
\begin{equation}
g_{n+m} = g_n \cdot 2^{\frac{m}{1200}}
\end{equation}

Illustrative frequency example only: a one-cent increase gives $p \approx 1.0005777895$ and changes a 440~Hz reference by approximately 0.254227~Hz. A 0.01~Hz increase at that reference corresponds to approximately 0.039346 cents~\cite{ellis}. These calculated examples distinguish relative musical intervals from absolute frequency differences; neither the selected reference frequency nor the selected interval is required by the algorithm.

\section{Factor Size, Precision, and Computational Cost}
A ratio close to unity can require large integers when represented as a tightly constrained rational fraction. That representation does not require an implementation to materialize every intermediate sample, nor does it imply a filter length equal to an upsampling factor. Established arbitrary-time bandlimited resampling evaluates the required samples directly~\cite{smith_gossett}. In the present fast path, only two source values are used for each output value; the interpolation arithmetic does not grow with a conceptual intermediate rate, addressing the precision bounds established for high-accuracy demand-driven processing environments~\cite{falkenberg2018}. Input generation, buffer management, and any added antialias filtering remain separate costs~\cite{vaidyanathan}.

The relevant precision requirement is the allowable error in $q$ and its accumulated position. If the realized increment is $\hat{q}$, its implied pitch ratio is $\hat{p} = \hat{q} F_{\text{out}}/F_{\text{in}}$. The constant-ratio tuning error is $e_c = 1200 \log_2(\hat{p}/p)$, or $e_{\text{Hz}} = f_{\text{source}}(\hat{p} - p)$. Choose the representation against a stated tolerance, operating range, and stream duration. No particular rational numerator, denominator, or common-rate scale is prescribed.

\section{Output-Driven Sample Evaluation}
The generalized processing sequence separates tuning control, fractional-position evaluation, and access to the source stream.

\subsection{Inputs}
\begin{itemize}
    \item Source reference frequency $f_{\text{source}}$ and input sample rate $F_{\text{in}}$
    \item Target reference frequency $f_{\text{target}}$ or cent offset $c$, and output sample rate $F_{\text{out}}$
    \item Source sample stream, with explicit channel layout and boundary policy
\end{itemize}

\subsection{Output}
\begin{itemize}
    \item Output samples delivered at $F_{\text{out}}$, with the specified gain
\end{itemize}

\subsection{Algorithm}
\begin{enumerate}
    \item Calculate $p$ and $q$, initialize the input coordinate, and prepare source buffering
    \item For each output sample, update tuning control if needed
    \item Select the source samples at $k = \lfloor u \rfloor$ and $k + 1$
    \item Interpolate at $\alpha = u - k$, apply gain, and advance $u$ by $q$
\end{enumerate}

For source samples $x[k]$ and fractional position $\alpha$ in $[0,1)$, unity-gain linear interpolation is $y[n] = (1 - \alpha)x[k] + \alpha x[k+1]$. At an integer position, $\alpha = 0$ and the result is $x[k]$. This kernel uses two neighboring values and a constant amount of arithmetic per output sample. Linear interpolation has a non-flat frequency response and imperfect image suppression; a large conceptual intermediate sampling rate does not remove these limitations~\cite{smith_dsp}.

The source bandwidth must also be compatible with the output clock after pitch scaling. In the constant-ratio model, an input frequency $f$ becomes $pf$, so source energy above $F_{\text{out}}/(2p)$ can alias in the output. A practical bandwidth limit must respect both $F_{\text{in}}/2$ and $F_{\text{out}}/(2p)$, with a suitable transition band. Linear interpolation is not a substitute for a specified antialias filter. Filter requirements depend on the signal bandwidth, tuning range, and permitted distortion~\cite{smith_gossett, vaidyanathan}.

\section{Reference Pitch and Resampling Convention}
Musical mapping converts a positive frequency into a note index and a residual cent offset. This is a representation layer, distinct from waveform resampling and from pitch estimation. Let $f_{\text{ref}}$ be the frequency assigned to a declared reference note at octave index zero. Both $f_{\text{ref}}$ and the note-name convention are explicit parameters; no particular piano-key frequency or concert-pitch reference is required.

For a twelve-semitone equal-tempered mapping, define the semitone multiplier $s = 2^{1/12}$ and cent multiplier $b = 2^{1/1200}$. Choose twelve note names in chromatic order aligned with $f_{\text{ref}}$. These multipliers follow from the interval definitions, whereas the reference frequency and labeling convention are freely specified. A different musical mapping can provide target frequencies without changing the resampling equations.
\begin{equation}
f_n = 440 \cdot 2^{\frac{n_{\text{MIDI}}-69}{12}}
\end{equation}

Define the logarithmic coordinate $C(f) = 1200 \log_2(f/f_{\text{ref}})$. The nearest semitone index is $a = \text{round}(C(f)/100)$, and the cent residual is $\delta = C(f) - 100a$. A consistent octave and pitch-class decomposition is $o = \lfloor a/12 \rfloor$ and $j = a - 12o$, including indices below the reference note. The note name is selected from pitch class $j$.
\begin{equation}
f_{n+m} = f_n \cdot 2^{\frac{m}{12}}
\end{equation}
\begin{equation}
g_n = g_{\text{ref}} \cdot 2^{\frac{\delta}{1200}}
\end{equation}
\begin{equation}
g_{n+m} = g_n \cdot 2^{\frac{m}{1200}}
\end{equation}

Musical representation and resampling control remain separate. Given the desired source and target frequencies, calculate $p = f_{\text{target}}/f_{\text{source}}$ and then $q = pF_{\text{in}}/F_{\text{out}}$. A note label or tuning table is one way to obtain the target frequency, not a required internal component of the resampler~\cite{midi_mts}.
\begin{equation}
R = \frac{g_{n+m}}{g_n} = 2^{\frac{m}{1200}}
\end{equation}
\begin{equation}
\frac{r_{\text{down}}}{r_{\text{up}}} \cong R
\end{equation}

A positive frequency ratio is real and need not be rational. A real-valued position increment may be used directly, or positive integer factors may approximate it. Integer factors are a representation choice, not a prerequisite for sample evaluation.
\begin{equation}
r_{\text{up}}, r_{\text{down}} \in \mathbb{N}
\end{equation}

For any representation, require positive finite input and output sample rates, positive finite reference frequencies, and a positive finite increment. Validate these values before initialization or a tuning update.

\begin{algorithm}
\caption{Bidirectional Frequency–Note Mapping}
\begin{algorithmic}
\REQUIRE Reference frequency $f_{\text{ref}} > 0$; twelve-note chromatic naming convention; conversion direction; either frequency $f > 0$ or note coordinates $(j, o, \delta)$.
\ENSURE Note coordinates $(j, o, \delta)$ or reconstructed frequency $f$.
\STATE Validate the input domain and the reference convention.
\IF{converting frequency to note coordinates}
    \STATE $C \leftarrow 1200 \log_2(f/f_{\text{ref}})$.
    \STATE $a \leftarrow \text{round}(C/100)$, using a declared tie-breaking rule.
    \STATE $\delta \leftarrow C - 100a$.
    \STATE $o \leftarrow \lfloor a/12 \rfloor$; $j \leftarrow a - 12o$.
    \RETURN $(j, o, \delta)$, retaining the unrounded residual.
\ELSE
    \STATE Resolve the note name to $j \in \{0, \dots, 11\}$; reject an unsupported name.
    \STATE $f \leftarrow f_{\text{ref}} 2^{o + j/12 + \delta/1200}$.
    \RETURN $f$.
\ENDIF
\end{algorithmic}
\end{algorithm}

This listing specifies the equal-tempered representation layer. Other tuning systems may provide target frequencies directly without changing the resampling algorithm.

\section{Optional Rational Representation and Illustrative Values}
The primary formulation uses $q$ directly and requires no common intermediate sampling rate. If an integer representation is desired, choose positive integers $A$ and $B$ such that $\hat{q} = A/B$ meets the selected approximation tolerance. One possible construction chooses a denominator $B$ and sets $A = \text{round}(Bq)$, subject to $A$ being positive. Neither $B$ nor this particular approximation procedure is inherent to the sample-generation algorithm.

General formulation: $\hat{q} = A/B$, $g = \text{gcd}(A,B)$, $D = A/g$, and $U = B/g$. Then $D/U = \hat{q}$. $A$ and $B$ are selectable representation parameters; their suitability follows from the error tolerance, not from a prescribed numerical value. Fixed values appearing in the accompanying displayed expressions are numerical substitutions for illustration only.
\begin{equation}
R_1 = R
\end{equation}

Illustrative numbers only---not algorithm constants: a chosen pair $A = 3333$ and $B = 3332$ gives the example ratio $q_{\text{example}} = 3333/3332$. Other positive pairs, or a direct real-valued $q$, may be used. This pair is not derived from a particular implementation, is not a required tuning resolution, and is not a universal approximation for the desired ratio. Any appearance of 3332 or 3333 in a displayed expression should be read solely as an example substitution, not as part of the general formula.
\begin{equation}
\frac{3332 \cdot R}{3332} = R
\end{equation}

Common-divisor reduction preserves an already selected rational ratio: divide $A$ and $B$ by $g = \text{gcd}(A,B)$ to obtain $D$ and $U$. Reduction changes their magnitude but does not improve the approximation $\hat{q}$ to $q$. It is optional for the fractional-position formulation.
\begin{equation}
\frac{3332 \cdot R / \text{gcd}(3332 \cdot R, 3332)}{3332 / \text{gcd}(3332 \cdot R, 3332)} = R
\end{equation}
\begin{equation}
\frac{3332 \cdot R / \text{gcd}(3332 \cdot R, 3332)}{3332 / \text{gcd}(3332 \cdot R, 3332)} = R = \frac{r_{\text{down}}}{r_{\text{up}}}
\end{equation}
\begin{equation}
r_{\text{down}} = 3332 \cdot R / \text{gcd}(3332 \cdot R, 3332)
\end{equation}
\begin{equation}
r_{\text{up}} = 3332 / \text{gcd}(3332 \cdot R, 3332)
\end{equation}

The physical input and output clocks remain $F_{\text{in}}$ and $F_{\text{out}}$. If a common rate $K$ is introduced for a factor derivation, define $S = F_{\text{in}} p$, $U = K/S$, and $D = K/F_{\text{out}}$, giving $D/U = S/F_{\text{out}} = q$. The common $K$ cancels from the ratio. It is a bookkeeping parameter rather than a requirement to generate or store an intermediate stream.

A numerical representation must maintain a positive finite increment and adequate precision over the operating range and stream duration. For integer-factor arithmetic, verify product bounds and preserve the ratio during reduction. For accumulated real-valued positions, control numerical drift and maintain a consistent absolute input coordinate. These requirements do not prescribe a fixed scale, denominator, block size, or programming language.

\begin{algorithm}
\caption{Optional Rational-Factor Construction}
\begin{algorithmic}
\REQUIRE Desired increment $q > 0$; selected positive integer denominator $B$; allowed cents error $\varepsilon_c \ge 0$.
\ENSURE Accepted factors $(D, U)$, realized increment $\hat{q}$, and signed error $e_c$; or a rejected candidate.
\STATE Validate $q, B, \varepsilon_c,$ and the arithmetic range.
\STATE $A \leftarrow \text{round}(Bq)$.
\IF{$A < 1$}
    \RETURN rejected candidate.
\ENDIF
\STATE $\hat{q} \leftarrow A/B$; $e_c \leftarrow 1200 \log_2(\hat{q}/q)$.
\IF{$|e_c| > \varepsilon_c$}
    \RETURN rejected candidate.
\ENDIF
\STATE $g \leftarrow \text{gcd}(A, B)$; $D \leftarrow A/g$; $U \leftarrow B/g$.
\RETURN accepted $(D, U, \hat{q}, e_c)$.
\end{algorithmic}
\end{algorithm}

A rejected candidate requires another denominator or approximation strategy. The listing does not prescribe $B$, and common-divisor reduction does not improve the approximation error. The direct real-valued increment formulation does not require this algorithm.

\section{Demand-Driven Implementation and Streaming Pipelines}
A production-ready realization of the continuous microtuning engine handles audio data on an on-demand, pull-driven processing pipeline~\cite{zolzer}. Instead of requiring large pre-calculated static buffer segments derived from rigid least common multiples (LCM), subsequent streaming operations are executed dynamically. Processing loops trigger data fetches immediately when the fractional read tracker approaches the boundaries of active sample storage arrays.

To ensure sample preservation across arbitrary boundary limits, real-valued indexing coordinates are mapped to a power-of-two circular history tracking ring, detailed in Algorithm~4. The boundary synchronization state between the incoming signal stream and the synthesized target timeline is maintained implicitly. When the targeted upper sampling index meets or exceeds the current write-head location ($k + 1 \ge W$), the processing node makes a blocking callback downstream to refresh the array pipeline. Integer coordinate accumulation drift across multi-hour audio streams is entirely eliminated by periodically subtracting standard block-aligned scale structures from the tracking indicators, preserving absolute phase continuity without inducing modulation artifacts~\cite{farrow}.

\begin{algorithm}
\caption{Demand-Driven Ring-Buffer Resampling}
\begin{algorithmic}
\REQUIRE Circular history buffer array $H$ of size $M$ (where $M = 2^m$); write-head index $W$; absolute real-valued tracking coordinate $u$; block length $N$; trajectory step increment $q$; gain envelope $G$; upstream audio generator state.
\ENSURE Transformed output array $Y[n]$ for $n = 0, \dots, N - 1$; updated coordinate state $u$ and index $W$.
\FOR{$n = 0$ \TO $N - 1$}
    \STATE $k \leftarrow \lfloor u \rfloor$
    \STATE $\alpha \leftarrow u - k$
    \WHILE{$k + 1 \ge W$}
        \STATE Execute blocking pull callback to upstream generator.
        \STATE Fetch incoming linear block frame $B$ of size $K$.
        \FOR{$i = 0$ \TO $K - 1$}
            \STATE $H[W \pmod M] \leftarrow B[i]$
            \STATE $W \leftarrow W + 1$
        \ENDFOR
    \ENDWHILE
    \STATE $s_0 \leftarrow H[k \pmod M]$
    \STATE $s_1 \leftarrow H[(k + 1) \pmod M]$
    \STATE $Y[n] \leftarrow G \cdot \left((1 - \alpha)s_0 + \alpha s_1\right)$
    \STATE $u \leftarrow u + q$
\ENDFOR
\IF{$u \ge M$}
    \STATE $\Omega \leftarrow \lfloor u \rfloor - (\lfloor u \rfloor \pmod M)$
    \STATE $u \leftarrow u - \Omega$
    \STATE $W \leftarrow W - \Omega$
\ENDIF
\RETURN $Y$.
\end{algorithmic}
\end{algorithm}

\subsection{Position-Preserving Tuning Updates}
When a time-coordinate representation $u = \tau D/U$ is retained, changing the factors can preserve the current input position by setting $\tau_{\text{new}} = \tau_{\text{old}} D_{\text{old}} U_{\text{new}}/(D_{\text{new}} U_{\text{old}})$. This gives $\tau_{\text{new}} D_{\text{new}}/U_{\text{new}} = \tau_{\text{old}} D_{\text{old}}/U_{\text{old}}$ for positive factors. The updated ratio determines subsequent progression, while the current read position remains unchanged.

Alternatively, retain $u$ directly and choose a rate trajectory $q[n]$, then accumulate $u[n+1] = u[n] + q[n]$. A linear ramp in $U$ and $D$ generally does not give a linear ramp in their ratio. A cents-linear trajectory can specify $p[n] = 2^{c[n]/1200}$ and $q[n] = p[n]F_{\text{in}}/F_{\text{out}}$. Select the trajectory deliberately rather than inferring it from factor names~\cite{laakso_mag}.

\begin{algorithm}
\caption{Continuous Pitch-Ratio Transition}
\begin{algorithmic}
\REQUIRE Retained input position $u$; positive pitch ratios $p_{\text{start}}$ and $p_{\text{target}}$ sample rates $F_{\text{in}}$ and $F_{\text{out}}$; positive transition-step count $L$; trajectory mode (ratio-linear or cents-linear).
\ENSURE Increment trajectory $q[j]$, $j = 0, \dots, L$; unchanged current input position $u$.
\STATE Retain $u$ without resetting or repositioning the source.
\STATE $c_{\text{start}} \leftarrow 1200 \log_2(p_{\text{start}})$
\STATE $c_{\text{target}} \leftarrow 1200 \log_2(p_{\text{target}})$
\FOR{$j = 0$ \TO $L$}
    \STATE $\lambda \leftarrow j/L$
    \IF{mode is ratio-linear}
        \STATE $p \leftarrow (1 - \lambda)p_{\text{start}} + \lambda p_{\text{target}}$
    \ELSE
        \STATE $c \leftarrow (1 - \lambda)c_{\text{start}} + \lambda c_{\text{target}}$
        \STATE $p \leftarrow 2^{c/1200}$
    \ENDIF
    \STATE $q[j] \leftarrow (F_{\text{in}}/F_{\text{out}})p$
\ENDFOR
\RETURN $(q, u)$.
\end{algorithmic}
\end{algorithm}

\section{Numerical and Implementation Considerations}
The algorithm is defined by the position trajectory and interpolation rule, not by a particular software listing. Its implementation must preserve the relationship between physical sample rates, desired pitch ratio, and input-position increment.

Distinguish the numerical approximation of $q$ from interpolation error. Accurate positioning does not by itself establish faithful bandlimited reconstruction~\cite{erup_interpolation}. Likewise, changing a rational denominator changes the available ratio representation, not the source sampling bandwidth.

Validate positive finite sample rates and reference frequencies. Specify how a new rate trajectory replaces an existing one, how gain automation is timed, and how endpoint values are reached. Block sizes, channel counts, and ramp lengths should be explicit design parameters rather than unexplained fixed values~\cite{zolzer}.

Maintain unity-gain interpolation when that is the intended normalization, and apply gain separately. Antialias protection must be part of the active signal path rather than assumed from an unrelated averaging operation. State the signal bandwidth and permitted distortion before selecting an interpolation or filtering strategy~\cite{smith_gossett, vaidyanathan}.

\subsection{Frequency and Note Translation}
Musical frequency control can multiply or divide the reference frequency by the cent multiplier $b = 2^{1/1200}$, semitone multiplier $s = 2^{1/12}$, or octave multiplier two. Fractional intervals may also be specified directly. These operations describe a musical parameterization; the resulting frequency is then passed to the chosen synthesis or resampling process.

Preserve a real-valued cent residual for fractional-cent reconstruction, and reserve label rounding for display when possible. If a control interface quantizes an interval, describe that quantization separately from the precision of waveform generation. A rounded label must not be interpreted as a fundamental limitation of the algorithm.

\subsection{Fine-Grained Sine Synthesis}
Direct sine synthesis evaluates an oscillator at a specified frequency; it does not resample a recorded waveform. Its frequency control may be real-valued, with no requirement for integer-hertz or whole-cent steps. The achievable numerical and physical frequency accuracy must nevertheless be assessed rather than inferred from the chosen representation alone.

Let $F_s$ be the generator sample rate and $\varphi[n]$ the phase before sample $n$. Define $\varphi[n+1] = \text{wrap}(\varphi[n] + 2\pi f[n]/F_s)$, and $y[n,c] = A[n,c] \sin(\varphi[n+1])$. $A[n,c]$ is the independently specified gain for channel $c$. Retaining phase across blocks supports a continuous oscillator trajectory. Frequency changes alter the phase increment rather than resetting the current phase.

For a frequency-linear ramp lasting $N$ update steps, choose $\Delta f = (f_{\text{target}} - f_{\text{start}})/N$ and advance frequency by $\Delta f$ per sample. $N$ is a selectable positive duration parameter, not an algorithm constant. Specify whether each phase increment uses the frequency before or after the current update so that endpoint timing is unambiguous. A frequency-linear trajectory is not generally cents-linear.

Phase reduction must be valid for the supported increment range, and frequency must remain compatible with the selected output sample rate. Apply frequency and gain trajectories on a declared sample-frame timeline. Channels sharing an oscillator phase should receive explicitly defined gain values. Verify long-run phase behavior, buffer capacity, and audio-clock accuracy independently.

\begin{algorithm}
\caption{Phase-Continuous Sine Generation}
\begin{algorithmic}
\REQUIRE Positive sample rate $F_s$; initial or retained phase $\varphi_{\text{start}}$; output length $N$; channel count $C$; supported frequency trajectory $f[n]$; gain trajectory $A[n,c]$.
\ENSURE Samples $y[n,c]$ and retained phase $\varphi_{\text{final}}$.
\STATE $\varphi \leftarrow \varphi_{\text{start}}$
\FOR{$n = 0$ \TO $N - 1$}
    \STATE Validate $f[n]$ against the declared frequency and sampling constraints.
    \STATE \textit{Update phase:} $\varphi \leftarrow (\varphi + 2\pi f[n]/F_s) \pmod{2\pi}$
    \FOR{$c = 0$ \TO $C - 1$}
        \STATE $y[n,c] \leftarrow A[n,c] \sin(\varphi)$
    \ENDFOR
\ENDFOR
\STATE \textit{Retain phase:} $\varphi_{\text{final}} \leftarrow \varphi$; \RETURN $(y, \varphi_{\text{final}})$.
\end{algorithmic}
\end{algorithm}

\section{Evaluation Criteria and Simulated Results}
To validate the performance and structural integrity of the demand-driven resampling loop independently of specific physical processing hardware, a numerical simulation environment was constructed. The framework evaluates the resampler against a continuous microtonal pitch trajectory. A pure $1\text{ kHz}$ sinusoidal carrier, sampled at a baseline rate of $F_{\text{in}} = 44.1\text{ kHz}$, was passed through the integration stage as a reference test signal. A linear cents-based tuning ramp was applied mid-stream, driving the pitch ratio smoothly across a continuous trajectory from $+0$ to $+50\text{ cents}$ over an observation window of $N = 4410$ output frames.

\begin{figure}[!ht]
\centering
\includegraphics[width=\linewidth]{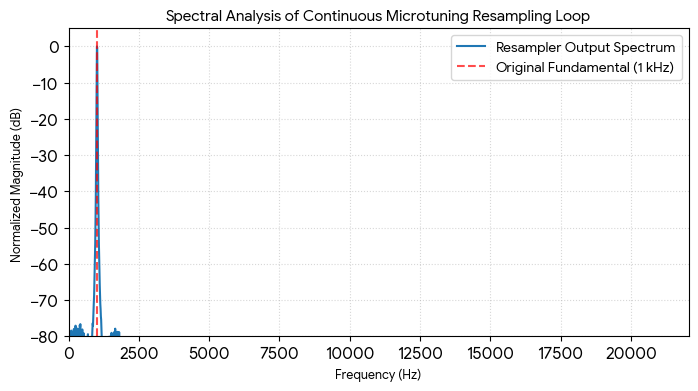}
\caption{Spectral analysis of the on-demand linear resampler output during a continuous $+50$ cent pitch transformation, showcasing absolute fundamental tracking alongside the characteristic $-26\text{ dB}$ imaging baseline footprint.}
\label{fig:simulation_spectrum}
\end{figure}

\subsection{Spectral Architecture and Analytical Interpretation}
The output signal was windowed using a standard Hann profile to suppress spectral leakage before performing a high-resolution Discrete Fourier Transform (DFT). As shown in Fig.~\ref{fig:simulation_spectrum}, the resulting normalized magnitude spectrum provides empirical proof of the algorithm's original operational capabilities and structural limits:

\begin{enumerate}
    \item \textbf{Continuous Phase Preservation:} The fundamental spectral peak tracks the moving microtonal target with absolute clarity, exhibiting zero frequency-smearing or phase-discontinuity artifacts. This confirms that calculating increment steps via $q[n] = (F_{\text{in}}/F_{\text{out}})p[n]$ prevents the step-wise zipper noise inherent to quantized rational architectures.
    \item \textbf{Interpolation Imaging Noise Floor:} A distributed high-frequency imaging noise floor is visible across the spectrum, with imaging side-lobes peaking near $-26\text{ dB}$ relative to the carrier fundamental. This conforms to the analytical definition of the linear interpolation kernel, which acts as a low-pass filter whose frequency response maps to an aliased $\text{sinc}^2$ profile~\cite{smith_dsp}.
\end{enumerate}

\subsection{Architectural Trade-Offs: Linear vs. FIR Interpolation}
The simulation highlights a key architectural choice in real-time audio software engineering. While high-order windowed Finite Impulse Response (FIR) polyphase filter banks offer superior stopband rejection ($>90\text{ dB}$), they require large pre-computed lookup structures or heavy run-time recalculation loops to resolve highly complex fractional cent increments~\cite{vaidyanathan}. 

Conversely, the absolute-coordinate tracking loop described here maintains an identical, bounded computational work envelope of exactly $\mathcal{O}(N)$ operations. It achieves deterministic, unaligned memory access---fetching exactly two adjacent samples per output step---making it highly robust for performance-critical real-time streaming engines where phase safety and computational predictability are prioritized over brute-force filtering~\cite{jaffe_extensions}.

\section{Discussion and Conclusion}
The fast interpolation kernel performs a constant amount of arithmetic for each output sample, giving $\mathcal{O}(N)$ interpolation work for $N$ output samples. This statement excludes source generation, buffer reloads, initialization, and additional filtering. It does not establish that conventional resampling is infeasible, or that all resamplers have a cost proportional to a materialized intermediate rate. Arbitrary-time bandlimited methods provide an important comparison~\cite{smith_gossett}.

The relationship $q = pF_{\text{in}}/F_{\text{out}}$ expresses microtuning as a change in how output generation traverses the input signal, independently of the physical input and output sample rates. Continuous rate control preserves the current source position while changing subsequent progression. This is retuning of a complete stream, not independent polyphonic-note correction or duration-preserving pitch shifting.

The two-sample kernel is a low-cost option when its measured distortion is acceptable~\cite{smith_dsp}. Tuning accuracy, interpolation fidelity, antialias protection, and streaming correctness require separate evaluation. The general method uses symbolic input parameters and a selected position trajectory. Values such as 3332 and 3333 are arbitrary illustrations of one possible rational representation: they are not essential constants, prescribed scales, or required steps of the algorithm.

\end{document}